\documentclass[%
 reprint,
 amsmath,amssymb,
 aps,
]{revtex4-1}

\usepackage{graphicx}
\usepackage{dcolumn}
\usepackage{bm}
\usepackage{graphicx}

\begin{document}

\preprint{APS/123-QED}

\title{Direct transfer of pump amplitude to parametric down-converted photons}

\author{Ali Anwar$^a$}
\author{Pravin Vaity$^{a,b}$}
\author{Chithrabhanu Perumangatt$^a$}
\author{R. P. Singh$^a$}
\affiliation{$^a$Physical Research Laboratory, Navarangpura, Ahmedabad, India-380009.}
\affiliation{$^b$Laboratory for Microactuators, IMTEK, University of Freiburg, Germany.}




\date{\today}

\begin{abstract}
We numerically and experimentally show that all photons generated by spontaneous parametric down-conversion (SPDC) follow a transverse amplitude similar to that of the pump. This amplitude transfer from pump to SPDC is revealed in the Fourier image plane of the down-converted photons restricted by an aperture. We also observe a considerable shift of the image plane from the actual Fourier plane, when size of the aperture is gradually increased. The shift of the Fourier image of down-converted photons affects the quality of spatial mode-based projection in various quantum correlation experiments with parametric down-converted photon pairs. The results may be useful in applications of down-converted photons for quantum imaging and quantum communication. 
\end{abstract}

\pacs{Valid PACS appear here}
\maketitle


Spontaneous parametric down-conversion (SPDC), one of the most common non-linear optical process, is of great significance in the field of quantum optics and quantum information \cite{kok}. In this second order non-linear optical process, a pump photon of higher energy, on interaction with a non-linear medium, gets down-converted to two lower energy photons generally called as signal and idler \cite{burnham}. SPDC is governed by conservation laws in energy and momentum where the later is called as phase-matching. Depending on the type of phase matching, parametric down-converted photons will have different geometrical distribution in space \cite{boydbook}. The process is very commonly used to generate heralded single photons by conditioning the pair photon \cite{castelletto} and their non-classical photon statistics has been investigated \cite{hockel,nogueiraprl}. The photon pairs generated in SPDC are correlated in space \cite{grayson}, time \cite{hong}, frequency \cite{baek}, polarization \cite{shih,kwiat} and orbital angular momentum \cite{mair}. Hence its been used as a source of entangled photon pairs in applications of quantum imaging and quantum information.

Heralded single photos produced by SPDC has been studied for their spatial and temporal coherence properties \cite{joobeur} is utilized in many fields like quantum imaging \cite{dangelo}. However, the individual signal or idler photons are incoherent in the absence of heralding. Thus in imaging, the direct image produced by signal or idler will not reveal anything about the object, while the heralded photons give better images. The two-photon correlation studies in SPDC has been carried out for various structured light pump beams \cite{arlt,walbornhg,mclaren}. In such cases, the two photon spatial modes obey the selection rules, giving rise to modal entanglement \cite{walbornoam,walbornhg,romero}. This can be measured by projecting the biphoton state to different modes using phase flattening techniques\cite{qassim}.

The angular spectrum (AS) and the conditional angular spectrum (CAS) of SPDC photons are highly dependent on the angular spectrum of the pump beam and crystal parameters \cite{ramirez,shashi}. It has been experimentally shown that the angular spectrum of the pump beam gets transferred to the twin photons generated in SPDC \cite{monken}. For example, it was shown that the amplitude as well as the helical phase of an optical vortex pump is transferred to the SPDC heralded single photon \cite{vicuna}. The helical phase of the single photon was verified by methods like triangular aperture diffraction \cite{vicuna} and interferometry \cite{galvez}. It was also observed that the SPDC heralded single photons generated with Bessel-Guassian pump show non-diffracting behaviour over a longer distance \cite{cruz}. So, images of high contrast and resolution can be obtained at any distance from the light source by using heralded single photons \cite{bennink}. Due to the incoherence of individual parametric down-converted photons, the angular spectrum of SPDC does not evidently show the signature of spatial properties of the pump.

Here, we observe that the transverse amplitude distribution of signal or idler beam mimics the pump amplitude distribution. We introduce an iris aperture on one arm of a non-collinear, degenerate SPDC and observe its image on the Fourier plane using a lens. We observe the same image on every part of the SPDC ring selected by an iris aperture. The size of the image formed increases when the aperture is moved away from the crystal. When the size of the aperture increases, the Fourier plane gets shifted from the actual '$2f$' Fourier plane. We perform phase measurement on the signal beam using phase flattening. We observe that though the signal beam mimics the transverse amplitude distribution of the pump, it does not follow the phase distribution of the pump.  

First we model the SPDC process for which the signal (or idler) photons are selected by an aperture  and image using a lens while idler (or signal) are undetected or discarded. In the perturbative treatment of spontaneous parametric down-conversion process, interaction of pump ($p$), signal ($s$) and idler ($i$) modes in a medium (Non-linear $\chi^{(2)}$ crystal) is represented by an interaction Hamiltonian $\mathcal{H}_I$. The initial state is a vacuum state $\vert0\rangle_s\vert0\rangle_i$, therefore the output state of SPDC is approximated as

\begin{equation}
\vert\Phi\rangle=\left(1-\frac{i}{\hbar}\int_{0}^{\tau}\mathcal{H}_I(t)dt\right)\vert0\rangle_s\vert0\rangle_i
\label{biphotonstate}
\end{equation}

The biphoton mode function of the generated twin photons in transverse momentum coordinates ($\mathbf{k}$) is obtained as
\begin{equation}
\Phi(\mathbf{k}^{\perp})={}_s\langle\mathbf{k}^{\perp}\vert_i\langle-\mathbf{k}^{\perp}\vert\Phi\rangle
\label{biphotonmode}
\end{equation}
where $\mathbf{k}^{\perp}$ and $-\mathbf{k}^{\perp}$ represents the transverse position of momentum coordinates of signal and idler respectively in the image plane. On simplification, the biphoton mode function in transverse momentum coordinates is given by
\begin{equation}
\Phi(\mathbf{k}^{\perp})=E_0(\mathbf{k}_p^{\perp})\text{sinc}\left(\dfrac{\Delta kL}{2}\right)\exp\left(i\dfrac{\Delta kL}{2}\right)
\label{biphotonmodefn}
\end{equation}
where $E_0(\mathbf{k}_p^{\perp})$ represents the pump transverse amplitude distribution, $\mathbf{k}_p^{\perp}$ is the angular coordinates of the pump, $\Delta k$ is the longitudinal phase mismatch, and $L$ is the thickness of the crystal. The exponential factor in the Eqn. (\ref{biphotonmodefn}) is a global phase term. Here, Laguerre Gaussian ($LG$) modes representing optical vortices are considered as pump spatial distribution. A typical $LG$ mode in cylindrical coordinates ($\rho,\phi$) is given by \cite{flossmann}
\begin{equation}
LG_p^l(\rho,\phi)=\frac{1}{w_p}\sqrt{\frac{2p!}{\pi (p+\vert l\vert)!}}\left(\frac{\sqrt{2}\rho}{w_p}\right)e^{-\frac{\rho^2}{w_p^2}}L_p^{\vert l\vert}\left(\frac{2\rho^2}{w_p^2}\right)e^{-il\phi}
\end{equation}
where $w_p$ is the pump beam waist at the crystal plane, $L_p^{\vert l\vert}()$ is the associated Laguerre polynomial in radial ($p$) and azimuthal ($l$) indices. $l$ indicates the OAM, $l\hbar$ carried by the beam. Different coaxial superposition of vortices is obtained by the addition and subtraction of the $LG$ modes. The angular spectrum of the down-converted photons is given by
\begin{equation}
R_s(\mathbf{k}^{\perp})=\int d(-\mathbf{k}^{\perp})\vert\Phi(\mathbf{k}^{\perp})\vert^2
\label{SPDCAS}
\end{equation}

After traversing a distance $z_0$ from the crystal, the SPDC photons are passed through an aperture. So, the resultant biphoton mode function in Fourier space by taking the Fourier transform of the product
\begin{equation}
\Phi_{(ap)}(\mathbf{k}^{\perp})= \mathcal{FT}\left[O\left(\vert \mathbf{k}^{\perp}\vert,-\frac{z_0}{k}\right)A(\mathbf{k}^{\perp})\Phi(\mathbf{k}^{\perp})\right]
\label{SPDCdistrbn_ap}
\end{equation}
where $A(\mathbf{k}^{\perp})$ is the aperture function in spatial frequency coordinates. $O(\vert\mathbf{k}^{\perp}\vert,-\frac{z_0}{k})$ is the optical transfer function (OTF) for the SPDC propagated from crystal plane to the aperture plane. Here, $k$ is the magnitude of the wave-vector of the signal/idler. A general expression of OTF of a beam with transverse wave-vector $\mathbf{k}^{\perp}$ and propagating a distance $b$ is given by \cite{williams}
\begin{equation}
O\left(\vert \mathbf{k}^{\perp}\vert, b\right)=\exp\left(i\frac{b}{2}\vert \mathbf{k}^{\perp}\vert^2\right)
\label{OTF}
\end{equation}
The distribution of signal photons is obtained by taking the partial trace of the mode function over the coordinates of the idler, which is written as
\begin{equation}
T_s(\mathbf{k}^{\perp})=\int d(-\mathbf{k}^{\perp})\Phi_{(ap)}^*(\mathbf{k}^{\perp})\Phi_{(ap)}(\mathbf{k}^{\perp})
\label{SPDCdistrbn}
\end{equation}

By controlling the diameter of the circular aperture, the SPDC amplitude distribution becomes identical to that of the pump.

The experiment consists of a pump beam of wavelength $(405 \pm 2) nm$ from TOPTICA iBeam Smart diode laser, incident on a Type-I phase-matched $\beta$-Barium Borate crystal of thickness $5 mm$ and transverse dimensions of 6 mm$\times$6 mm with an optic axis oriented at 29.97$^\circ$ to the normal incidence. 
\begin{figure}[htbp]
\centering
\includegraphics[width=0.8\linewidth]{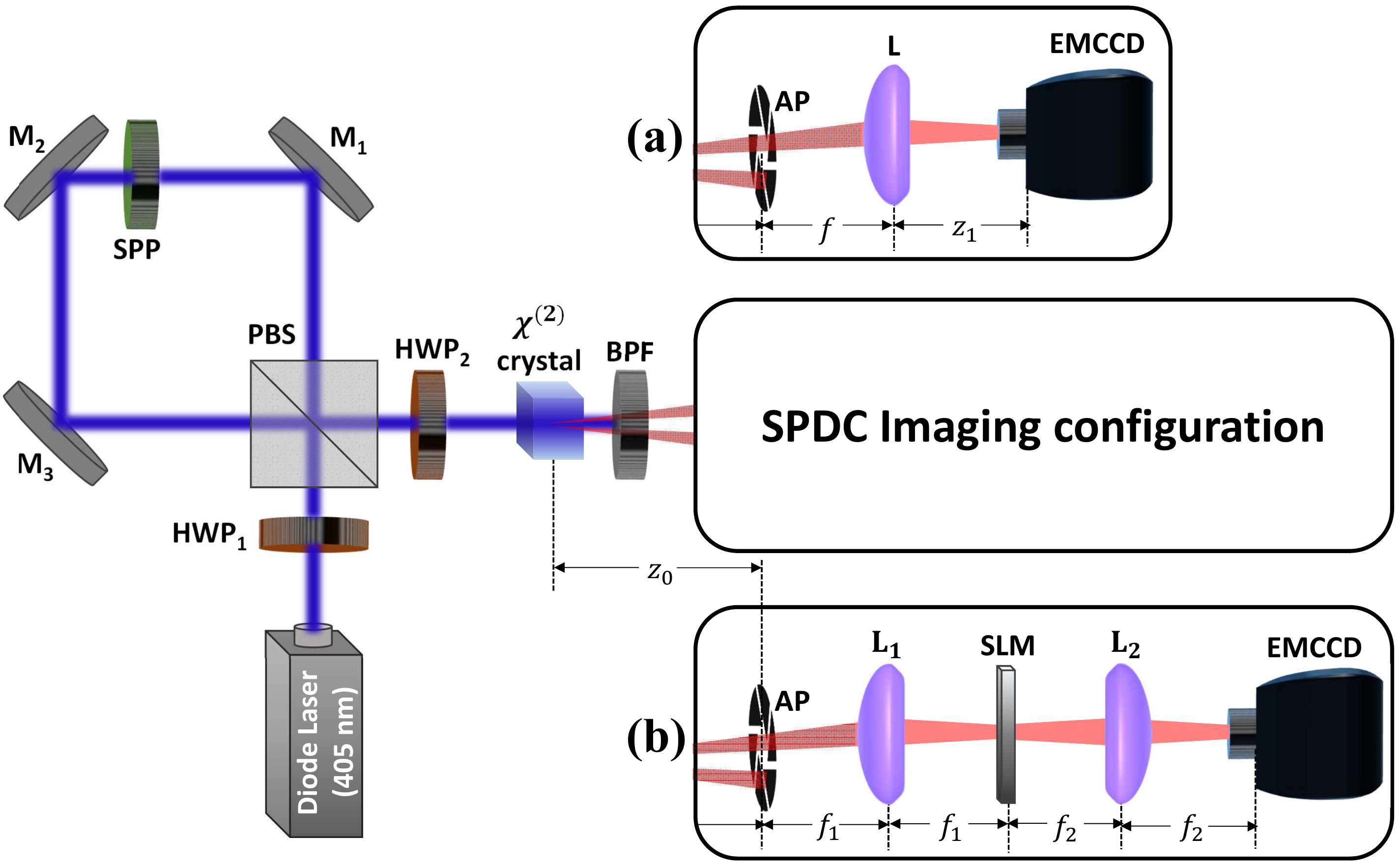}
\caption{Experimental setup for imaging parametric down-converted photons restricted by an iris aperture. (a) Fourier imaging configuration. (b) 'Phase-flattening' configuration. Here, the focal length of lenses L$_1$ and L$_2$ are $500 mm$ and $300 mm$ respectively. HWP$_1$, HWP$_2$ - Half-wave plate, PBS - Polarizing beam splitter, SPP - Spiral phase plate, M$_1$, M$_2$, M$_3$ - Mirrors, BPF - Band pass filter, AP - Iris aperture, L, L$_1$, L$_2$ - Plano-convex lenses, SLM - Spatial Light Modulator, EMCCD - Electron Multiplying CCD Camera.}
\label{expt_setup}
\end{figure}
Here, optical vortices of order 1, 2 \& 3 and their equal but opposite coaxial superposition are used as different pump spatial modes. To generate the vortex superposition, a modified polarizing Sagnac interferometer \cite{chithra,slussarenko} is set up before the crystal, as shown in Fig.\ref{expt_setup}. The first half-wave plate ($HWP_1$) is used to equalize the intensity of the counter-propagating beams inside the interferometer. A spiral phase plate ($SPP$) of desired order is introduced in the interferometer such that the counter-propagating beams acquire equal and opposite spiral phases before recombining at the polarizing beam splitter ($PBS$). In fact, the beam coming out of the interferometer is a vector vortex beam \cite{zhan,aadhi}. Based on the orientation of the second half-wave plate ($HWP_2$) kept before the crystal, crystal will down-convert only those modes whose polarization direction is along the optic axis of the crystal. The down-converted photons (signal \& idler) of wavelength 810 nm each (degenerate) are generated in a non-collinear fashion at diametrically opposite points of the SPDC ring. A bandpass filter (BPF) of central wavelength $(810 \pm 5) nm$ is used to block the pump beam after the crystal. The Fourier imaging configuration of SPDC photons through an iris aperture is shown in Fig.\ref{expt_setup}(a).

To show the effect of aperture on the spatial distribution of SPDC, the configuration shown in Fig.\ref{expt_setup}(a) is arranged. 
\begin{figure}[htbp]
\centering
\includegraphics[width=0.9\linewidth]{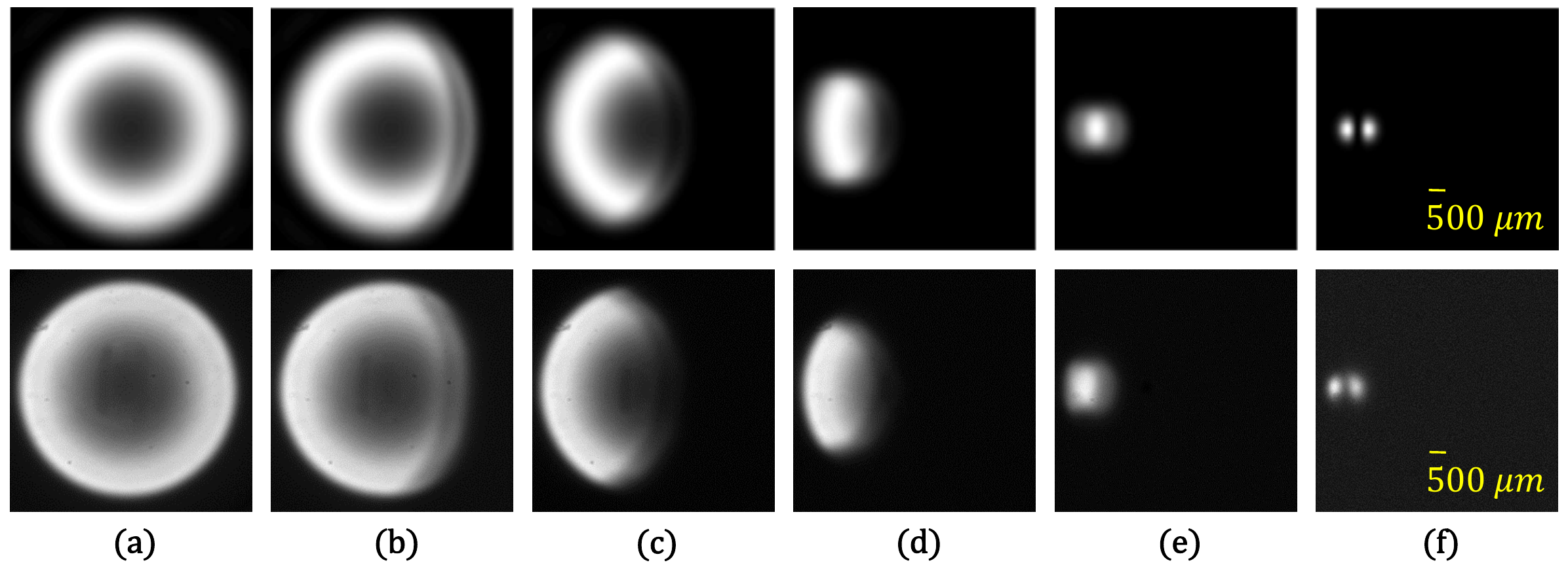}
\caption{Numerical (top row) and experimental (bottom row) angular spectrum of parametric down-converted photons at the Fourier plane of the aperture. On gradual closure of the aperture to minimum ((a)-(f)), the SPDC photons form a first order Hermite-Gaussian intensity distribution same as that of the pump beam.}
\label{close_aperture}
\end{figure}
An iris aperture (ID15, THORLABS) is kept on a portion of the SPDC intensity distribution at a distance $z_0=5cm$ from the crystal plane and a plano-convex lens of focal length $10cm$ is placed at a distance $f=10cm$ from the iris plane. For imaging the SPDC photons, an electron multiplying CCD (EMCCD) camera (Andor iXon3) of 512$\times$512 pixels with a pixel size of 16$\times$16 $\mu m^2$ is placed at a distance $z_1=10cm$ from the lens, which is supposed to be the Fourier plane of the aperture $AP$. First, the spatial distribution of parametric down-converted photons (SPDC annular ring) is imaged at the Fourier image plane of the iris. 
\begin{figure}[htbp]
\centering
\includegraphics[width=0.8\linewidth]{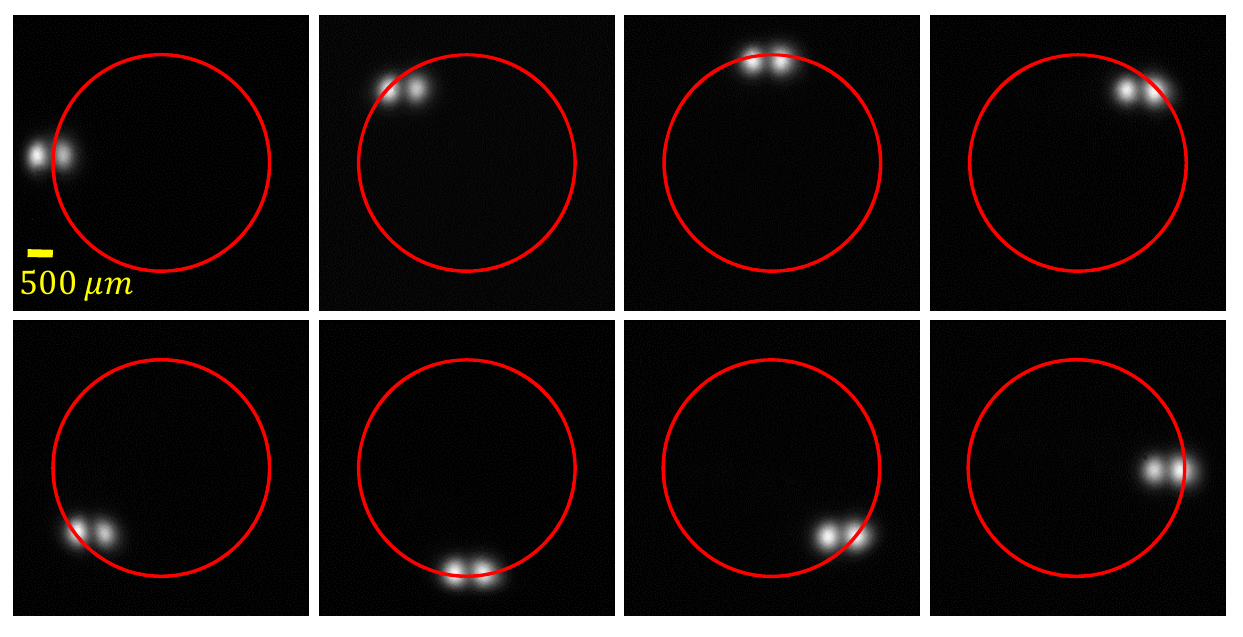}
\caption{Angular spectrum of parametric down-converted photons at the Fourier plane of the aperture placed at eight different portions on the SPDC annular ring. A red circle is drawn on each sub-figure to show the position of the image corresponding to the portion selected on the SPDC ring.}
\label{SPDC_allparts}
\end{figure}
Then, on selecting a portion of the ring, an intensity distribution identical to that of the pump is seen in the image plane. Here, the pump beam is a first order Hermite-Gaussian mode formed by the superposition of optical vortices of orders $+1$ \& $-1$ \cite{padgett}. Because the individual photons in SPDC are incoherent, the image formed at the Fourier plane of the aperture is almost free from irregularities due to aperture diffraction.
\begin{figure}[htbp]
\centering
\includegraphics[width=0.9\linewidth]{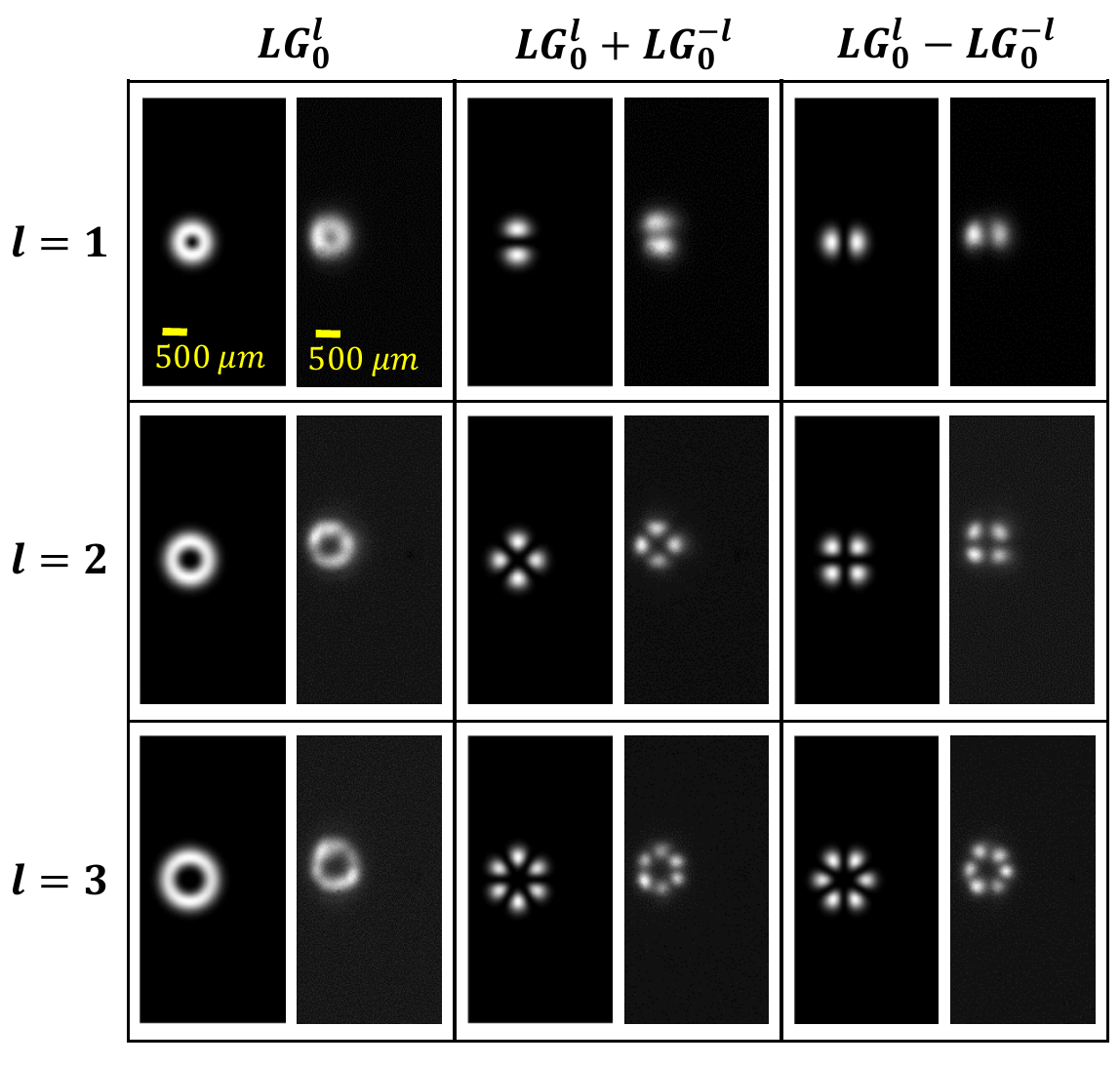}
\caption{Angular spectrum of parametric down-converted photons at the Fourier plane of the closed-to-minimum aperture when the non-linear crystal is pumped with optical vortices of orders 1, 2, 3 and their equal but oppositely charged coaxial superpositions, selected using $HWP_2$. Left part of each sub-figure shows the numerical image and the right part shows the experimental image.}
\label{diff_modes}
\end{figure}
Figure \ref{close_aperture} shows the images of the SPDC ring with gradual closure of the iris. Each image is recorded with the addition of 100 frames for an exposure time of 500 $\mu s$ each. The same intensity pattern is observed irrespective of the portion selected from the SPDC ring. Figure \ref{SPDC_allparts} shows first order HG mode formed at eight different parts of the SPDC ring annular distribution.  With the adjustment of the $HWP_2$ and the use of different spiral phase plates in the interferometer, the SPDC image is observed for different pump modes. Figure \ref{diff_modes} shows the numerical and experimental images of SPDC beam with minimal closure of iris for different pump polarization modes selected using $HWP_2$. Equation \ref{SPDCdistrbn_ap} is used to generate the numerical images with a circular aperture function.

Its well known that the SPDC biphoton mode has a similar phase profile that of the pump. Here the phase profile of the signal or idler photon modes is studied without the heralding. 
\begin{figure}[htbp]
\centering
\includegraphics[width=0.8\linewidth]{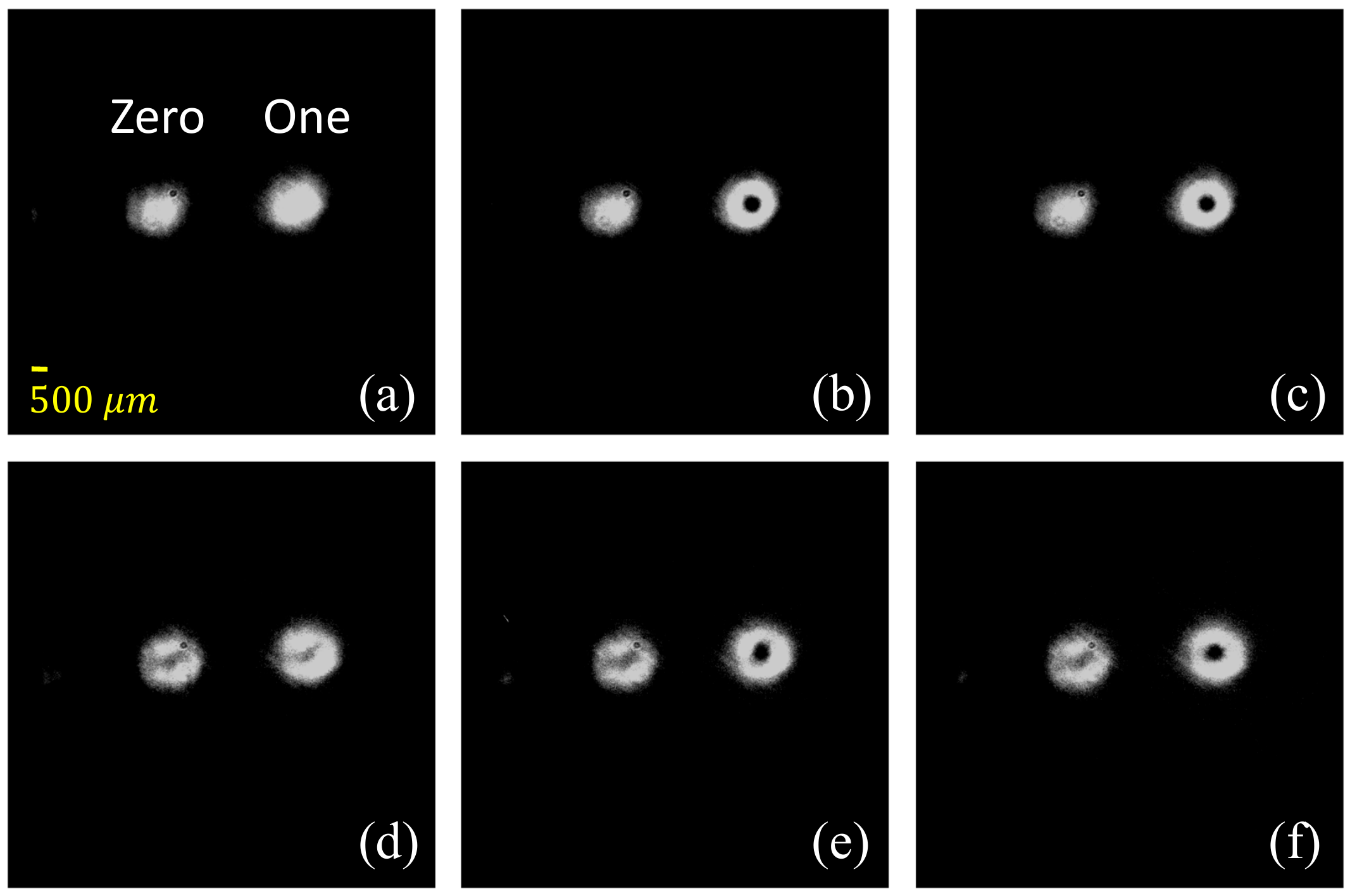}
\caption{(a)-(c) Far-field zeroth and first order diffraction pattern of SPDC pumped with (a)-(c) Gaussian  and (d)-(f) third order optical vortex, and projected onto the zeroth, $+3$ \& $-3$ order forked holograms respectively.}
\label{phaseflatten}
\end{figure}
A 'phase-flattening' method \cite{qassim} is used to verify whether the observed mode distribution has an azimuthal phase when pumped with a vortex beam. The Fourier plane of the aperture is projected onto the holograms imprinted on a spatial light modulator (SLM - Hamamtsu LCOS) and the far-field diffraction pattern is imaged (Fig.\ref{expt_setup}(b)). Figure \ref{phaseflatten}(a)-(c) show the diffraction patterns of the SPDC pumped with Gaussian, and projected onto the $0$, $+1$ and $-1$ order forked holograms respectively. When a vortex of order 3 is pumped and the intensity distribution of the aperture plane is projected onto a $+3$ as well as $-3$ forked holograms, a Gaussian-like shape is not observed in the first-order diffraction pattern, which must be observed if the projected mode contains an azimuthal phase of $\pm 6\pi$. This shows that the 'pump-mimicked' mode distribution of SPDC does not contain azimuthal phase corresponding to that of the pump. 
\begin{figure}[htbp]
\centering
\includegraphics[width=0.8\linewidth]{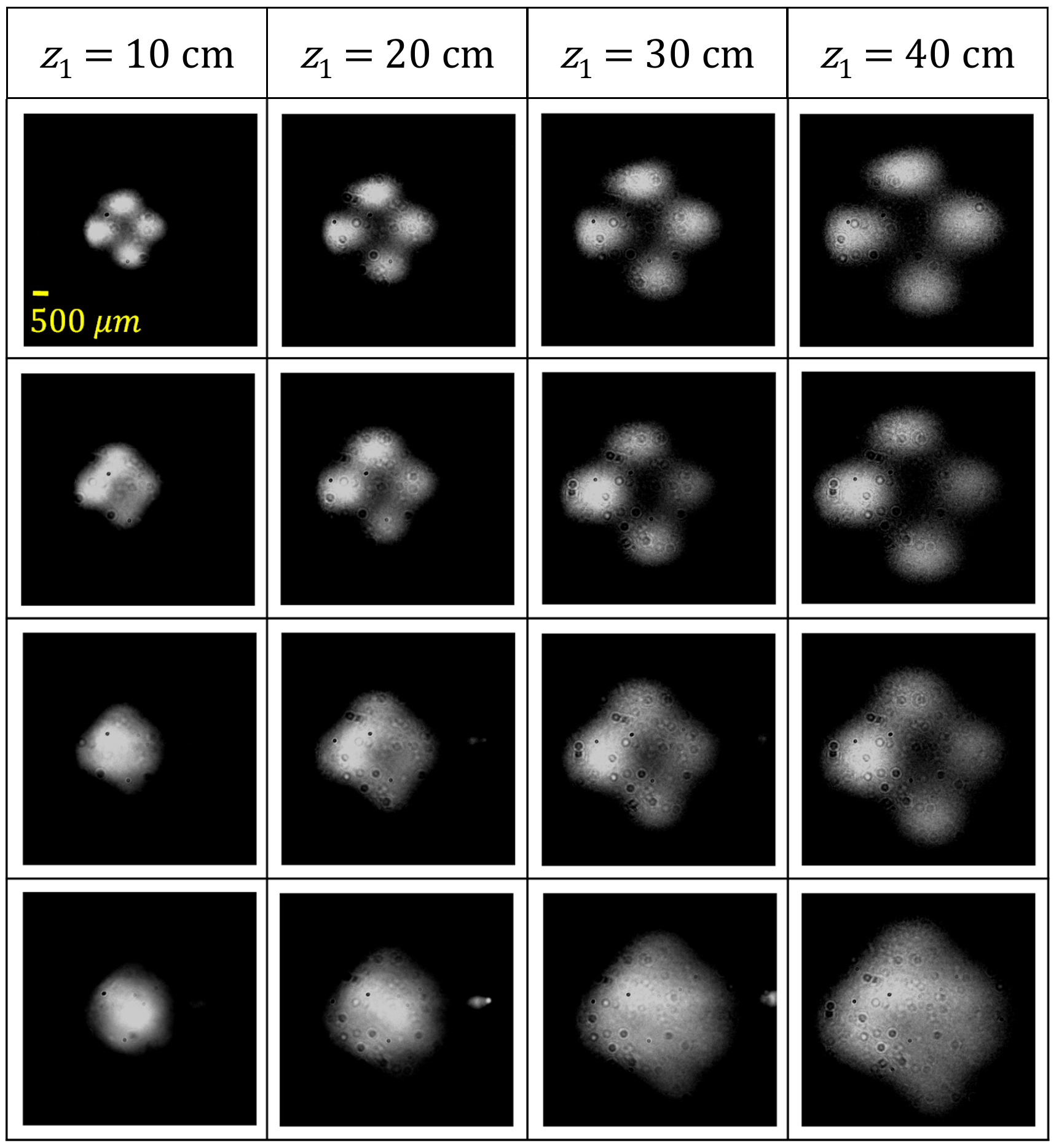}
\caption{Images of parametric down-converted photons at different $z_1$-distances for different iris aperture sizes with a superposition of optical vortices of orders $+2$ and $-2$ as the pump. The diameters of the apertures from top to bottom are $102.3\mu m$, $132.5\mu m$, $174.4\mu m$ and $246.4\mu m$ respectively.}
\label{ap_prop}
\end{figure} 
It is expected that the individual photons will not posses a particular azimuthal phase because they will be in a mixed OAM state while the two photons are entangled in OAM. However, the corresponding heralded single photon generated by SPDC process contain the azimuthal phase that has been well studied in the context of OAM entanglement \cite{galvez,mair}.

To study the shift of the image plane with the iris aperture size, images of the SPDC photons is recorded at different distances by moving the camera away (increasing $z_1$) for different aperture sizes, as shown in Fig. \ref{ap_prop}. Here, a superposition of optical vortices of orders $+2$ and $-2$ is used as the pump beam. The amplitude distribution of the pump is observed in SPDC photons around the distance $z_1=10cm$ for smallest aperture size that continues for longer distances beyond $z_1=10cm$, although the contrast become poorer. As we increased the aperture size, the plane of image formation shifts beyond the actual '$2f$' Fourier plane and the range of the distance in which the image is present was reduced, showing the image formation only at a particular plane for larger aperture sizes.

In conclusion, we have observed that all parametric down-converted photons follow the transverse amplitude profile as that of the pump. SPDC photon reveals the pump mode, when selected by a closed aperture and observed at the Fourier plane. With gradual increase in size of the aperture, the image plane is observed to be shifted from the actual $2f$ Fourier imaging distance. Further, we show that the individual photons does not reveal anything about the transverse phase profile of the pump. The results will be useful to improve the quality and efficiency of the biphoton mode coupling based on spatial mode projection for applications in multi-dimensional quantum information schemes.
\newline

\textbf{Aknowledgement}: The authors thank Dr. Vijay Kumar for the helpful discussions to carry out the work in this letter during his affiliation to Physical Research Laboratory (PRL), Ahmedabad, India. The
numerical computations were carried out using the Vikram-100 HPC cluster of PRL.

\end{document}